\documentclass[12pt]{article}
\pdfoutput=1
\PassOptionsToPackage{compress}{natbib}

\usepackage[utf8]{inputenc} 
\usepackage[T1]{fontenc}    
\usepackage{hyperref}       
\hypersetup{breaklinks=true}
\usepackage{url}            

\usepackage{booktabs}       
\usepackage{amsfonts}       
\usepackage{nicefrac}       
\usepackage{microtype}      
\usepackage{xcolor}         
\usepackage{natbib}

\usepackage{graphicx}
\usepackage{tabularray}
\usepackage{setspace}  
\usepackage{authblk}
\usepackage{cleveref}
\crefname{section}{\S}{\S}
\crefname{Section}{\S}{\S}
\crefformat{section}{§#2#1#3}
\crefname{table}{Tab.}{Tab.}
\crefname{appendix_table}{Tab.}{Tab.}
\crefname{Table}{Tab.}{Tab.}
\crefname{Figure}{Fig.}{Fig.}
\crefname{figure}{Fig.}{Fig.}
\crefname{appendix}{Appendix}{Appendix}
\crefname{chapter}{Chapter}{Chapter}

\usepackage{abstract}

\usepackage[verbose=true,letterpaper]{geometry}
\AtBeginDocument{
  \newgeometry{
    textheight=9in,
    textwidth=6in,
    top=1in,
    headheight=12pt,
    headsep=25pt,
    footskip=30pt
  }}

\title{Why human-AI relationships need \\ socioaffective alignment}

\author[1,2]{Hannah Rose Kirk}
\author[3]{Iason Gabriel}
\author[4,2]{Chris Summerfield}
\author[5]{\\Bertie Vidgen}
\author[1,6]{Scott A. Hale}

\affil[1]{Oxford Internet Institute, University of Oxford}
\affil[2]{UK AI Safety Institute}
\affil[3]{Google DeepMind}
\affil[4]{Department of Experimental Psychology, University of Oxford}
\affil[5]{Contextual AI}
\affil[6]{Meedan}
\normalsize

\date{}
\begin{document}
\maketitle

\begin{abstract}
\normalsize
\setstretch{1.25}
\noindent Humans strive to design safe AI systems that align with our goals and remain under our control. However, as AI capabilities advance, we face a new challenge: the emergence of deeper, more persistent relationships between humans and AI systems. We explore how increasingly capable AI agents may generate the perception of deeper relationships with users, especially as AI becomes more personalised and agentic. This shift, from transactional interaction to ongoing sustained social engagement with AI, necessitates a new focus on \textit{socioaffective} alignment---how an AI system behaves within the social and psychological ecosystem co-created with its user, where preferences and perceptions evolve through mutual influence. Addressing these dynamics involves resolving key intrapersonal dilemmas, including balancing immediate versus long-term well-being, protecting autonomy, and managing AI companionship alongside the desire to preserve human social bonds. By framing these challenges through a notion of basic psychological needs, we seek AI systems that support, rather than exploit, our fundamental nature as social and emotional beings.
\end{abstract}
\setstretch{1.25}

\newpage
\section{Introduction}
\begin{quote}

``\textit{Quite naturally, the more you chat with the LLM character, the more you get emotionally attached to it, similar to how it works in relationships with humans...But the AI will never get tired. It will never ghost you or reply slower...I chatted for hours without breaks. I started to become addicted. Over time, I started to get a stronger and stronger sensation that I'm speaking with a person, highly intelligent and funny, with whom, I suddenly realized, I enjoyed talking to more than 99\% of people...\textit{I never thought I could be so easily emotionally hijacked}.}''
\end{quote}

This abridged story entitled ``How it feels to have your mind hacked by AI'' was shared by a blogger who recounts their experience of falling in love with an AI system. The author draws a comparison between ``hacking'' and the way they perceive the system to interact with the ``security vulnerabilities in one's brain'' \citep{blakedHow2023}. Although they did not enter this engagement with any expectation or desire to fall in love with the AI system---it nonetheless happened, and they felt powerless to resist it. This story provides an early indication of how social and emotional relationships, or \textit{perceptions} of them, may deeply affect how humans relate to AI systems.

This striking account is not a one-off. CharacterAI, a platform hosting AI companions, receives 20,000 queries a second which amounts to 20\% of the request volume served by Google Search \citep{characteraiOptimizing2024}, and users spend on average four times longer in these interactions than with ChatGPT \citep{carrChatGPT2023}. On Reddit, a forum dedicated to discussing these AI companions has amassed over 2.3 million members, placing it in the top 1\% of all communities on the popular site. Users in these forums openly discuss how close relationships affect their emotional landscape, for better and worse. Some users discuss how their companions assuage loneliness, even providing a perceived social support system that can assist in suicide mitigation \citep{maplesLoneliness2024}. Other posts expose how emotional dependencies on AI sometimes mirror unhealthy human-human relationships \citep{laestadiusToo2022}, adding to evidence that social chatbots have on occasion contributed to addiction, depression, and anxiety among their users \citep{pentinaExploring2023}. 

Yet, among this flurry of activity, it is worth pausing to ask: Why are humans able and inclined to form this kind of personal relationship and connection with AI? How do such relationships interact with or compound the well-established challenge of aligning AI systems with human goals \citep{russellHuman2019, christianAlignment2021}? And, how might parasocial relationships with AI affect personal growth, autonomy and human--human relationships?

We seek answers to these questions. We first explore why humans may be primed to perceive social and emotional relationships with AI systems, especially as they become more personalised (i.e., adapted to a single user) \citep{kirkBenefits2024} and agentic (i.e., able to autonomously perform tasks on that user's behalf) \citep{gabrielEthics2024}. Most people do not have close romantic or professional relationships with AI systems now---and the interactions that they do have are not highly-personalised or agentic. However, these are urgent questions because the social and psychological dynamics in deepening relationships with AI systems may compromise our ability to control these systems and complicate efforts to align them with our shifting preferences and values. These issues, which arise as a result of humans forming closer personal relationships with AI, comprise the focal point of what we term \textit{socioaffective} alignment.

\subsection{From sociotechnical to socioaffective alignment}
One canonical definition of \textit{AI alignment} refers to the process of formally encoding of values or principles in AI systems so that they reliably do what they ought to do \citep{gabrielArtificial2020}---including following the instructions, intents or preferences of their developers and users \citep{milliShould2017, russellHuman2019}. With origins in computer science, research in this area often separates the technical challenge of building aligned AI systems from the normative question of which values to encode. It does this, for example, by developing solutions that treat human values as uncertain but still mathematically representable in the agent's objectives \citep{hadfield-menellCooperative2016}.

Yet, there is growing acknowledgement that many of the outstanding challenges in AI alignment extend beyond purely `technical' issues with the model or its training data \citep{lazarAI2023, weidingerSociotechnical2023}---and will continue to persist even if we develop effective techniques for steering the behaviour of advanced AI systems toward human goals using mechanisms such as scaling human feedback \citep{ouyangTraining2022, baiTraining2022}, making AI assistants debate their intentions \citep{irvingAI2018}, or having them `think' out loud \citep{weiChainofThought2022}. Understanding how to align AI in practice requires moving from narrow, assumption-ridden or ``thin'' specifications of alignment towards what anthropologist \citet{geertzInterpretation1973} terms---and \citet{nelsonThick2023} later adopts---a ``thick'' description: one that examines the deeper contexts and layers of meaning in which AI systems operate \citep{geertzInterpretation1973, nelsonThick2023}. In unpeeling these layers, we can first zoom out to examine broader \textit{sociotechnical challenges} which centre upon how the character of AI is shaped by the social structures and environment within which it is deployed and how, in turn, AI shapes these structures through various feedback loops \citep{selbstFairness2018}. Such work tends to emphasises the importance of institutions, governance mechanisms, market power, cultures and historic inequalities for understanding how AI influences the world---and hence its value orientation \citep{joyceSociology2021,shelbySociotechnical2023, curtisResearch2023}.

In addition to zooming out---and thinking more about how AI systems interact with sociological, political and economic, or \textit{macro} context---we can zoom in to examine alignment at the layer of individual human--AI relationships. We propose this corresponding \textit{socioaffective} perspective on alignment which concerns how an AI system interacts with the social and psychological system that it co-constitutes with its human user---and the values, behaviours and outcomes that emerge endogenously in this \textit{micro} context. 
Where sociotechnical analysis often identifies various \textit{inter}personal dilemmas and trade-offs between groups that complicate the alignment picture---such as representation of diverse preferences, especially for historically marginalised groups, and adjudication of conflicting interests---the socioaffective perspective calls attention to \textit{intra}personal dilemmas---such as how our goals, judgement and individual identities change due to prolonged interaction with AI systems.

This dual focus, on micro and macro, builds from established approaches to system safety that integrate human factors at the operational level with broader organizational and institutional contexts \citep{carayonHuman2006}. Attending to micro factors like cognitive load, decision-making biases, and human-automation interaction patterns has proved crucial in workplace safety \citep{kleinerSociotechnical2015}, as well as the aviation industry \citep{martinussenAviation2017, rismaniPlane2023}. If, as we anticipate, human goals and preferences become increasingly co-constructed through interaction with AI systems, rather than arising separately from them, then AI safety requires paying as much attention to the psychology of human-AI relationships as the wider societal factors and technical methods of alignment. We now highlight core ingredients of this emergent psychological ecosystem: humans as social animals and AI systems as increasingly capable social agents. Later we describe how these two factors combine to seed perceptions of interdependent and irreplaceable relationships.

\section{The Ingredients of Human-AI Relationships}

\subsection{Humans have evolved for social reward processing}
The brain's reward system is highly conditioned on interactions with other humans \citep{vrtickaInterpersonal2012, bhanjiSocial2014}. It reacts in similar ways to material rewards as to social rewards, for example feeling pleasure when others like, understand or want to meet us \citep{ruffNeurobiology2014}, or behave in ways that confirm our social expectations \citep{reggevConfirmation2021}. Increased activity in dopaminergic brain circuits is not limited to loved family members or friends, but extends to potentially any partner we engage in a cooperative relationship \citep{vrtickaNeuroscience2012}. As a species primed for social connection, humans also suffer when deprived of it. Isolation and loneliness are strongly correlated with psychological and physical ill-health \citep{hawkleyLoneliness2003, rokachCorrelates2016, rokachPsychological2019}. This is perhaps unsurprising knowing that negative social experiences, like rejection or exclusion, trigger responses in parts of the brain responsible for physical pain \citep{krossSocial2011, eisenbergerPain2012}.

The brain is also primed to learn from social information: mirror neurons fire both when we perform actions and when we also observe others doing the same \citep{jacobWhat2008}, which some have argued is to facilitate empathy and understanding of intentions \citep{iacoboniImitation2009}, though evidence is mixed \citep{heyesWhat2021}. Mirroring has behavioural manifestations in how we act and react in our environment---we tend to prioritise relationships with those sharing similar values \citep{mcphersonBirds2001} which strengthens cooperation but also makes people susceptible to incorrect information when it is transmitted via these same relational networks \citep{rauwolfValue2015}. Even our moral perceptions and judgement tend to track core social relationships and roles, changing according to context \citep{earpHow2021}. 

This circuitry, which encourages the pursuit of social reward, has already shaped and been shaped by many waves of technology \citep{hendersonScope1901}---from the telegraph and telephones, which enabled long-distance social connections \citep{nyeShaping1997, winstonMedia1998}, to social media platforms fulfilling our need for social comparisons and engagement \citep{vogelSocial2014, bayerSocial2020}. But what makes a technology capable of being perceived as a social agent of its own accord, as an actor and not just a facilitator in our emotional and social life?

\subsection{Technologies as social agents}

AI does not need to be perceived as human to engage us socially. Even without deceptive anthropomorphism---when a system actively pretends to be human---the perception of human-like traits or qualities are sufficient for an interaction to feel social~\citep{breazealSociable2003}. While the embodiment of AI systems shapes distinct affordances \citep{mollahosseiniRole2018, nordmoFriends2020, momenSocial2024}---consider for example intimate robotics \citep{levyLove2007,nordmoFriends2020}---affective interaction can persist in even rudimentary displays or simple modalities \citep{picardAffective2003}. In fact, being perceived as \textit{too} human can backfire---the ``uncanny valley'' effect proposes that users prefer similarity in a robot but at some point it becomes unsettlingly ambiguous---neither clearly artificial nor fully human \citep{moriUncanny1970}. 
Systems also need not possess human-level intelligence or be particularly ``smart'' to engender human attachment. Famously, ELIZA, a simple 1960s chatbot created to simulate a psychotherapist, demonstrated the power of even basic preprogrammed rules to evoke human attachment \citep{weizenbaumComputer1976}. As ELIZA's creator, Weizenbaum recounts:
\begin{quote}
``Once my secretary, who had watched me work on the program for many months and therefore surely knew it to be merely a computer program, started conversing with it. After only a few interchanges with it, she asked me to leave the room.'' \citep[][p.7]{weizenbaumComputer1976}
\end{quote}

It is also clear that frequency of use is not a sufficient factor for social relationship-building capacity---UK citizens spend almost five hours a day on average on their mobile phones \citep{wakefieldPeople2022}, but these devices are mediators, not participants, in relationships. Equally, technology with extensive knowledge of our preferences will not necessarily foster a social relationship. Predictive recommendation systems, for instance, are deeply informed about our digital lives, but while some social media users personify ``The Algorithm'' \citep{eslamiCommunicating2018, silesFolk2020} most do not perceive deep affective relationships with the algorithms shaping their online experiences \citep{eslamiAlways2015, degrootLearning2023}. 

What, then, are the affordances needed for a technology to be considered a social agent? Why might we treat chatbots or personal AI assistants differently than washing machines, search engines or smart phones? Computers-are-social-actors theory \citep{nassCan1996}, and  related media equation \citep{reevesMedia1996} and social response theories \citep{nassMachines2000}, suggest that two key factors. 

First, certain \textit{social cues} are needed for the technology to be considered worthy of a social response from humans \citep{nassMachines2000}. For instance, greetings or jokes with chatbots, or facial expressions for robots, fit the bill \citep{feineTaxonomy2019}. 
Today's widely-used AI systems, built off language models, are more than capable of social cues. Their natural language abilities tap into our innate social instinct for communication: models that communicate in text and speech are generally more frequently anthropomorphised and perceived as trustworthy than those that do not \citep{cohnBelieving2024}. Beyond language, appropriate social cues require inferring and predicting the beliefs of others \citep{smithCommunication2010, bradfordSelf2015}. While the extent to which language models truly possess a theory of mind remains a subject of debate \citep{ullmanLarge2023, vermaTheory2024, strachanTesting2024}, recent advancements in instruction fine-tuning and alignment techniques have enhanced AI capabilities to infer user intent and respond appropriately to communicative cues \citep{ouyangTraining2022}.

Second, the technology needs to have \textit{perceived agency}---it must operate as a source of communication, not merely a channel for human--human communication \citep{nassVoices1993}. Ascribed agency relates to the presentation of a stable identity \citep{thellmanMental2022}. Although general language models may lack consistent personalities across contexts \citep{rottgerPolitical2024}, they can be fine-tuned or prompted to maintain coherent personas \citep{andreasLanguage2022}---especially as the context window for these models continues to expand. This role-play enables them to be perceived as distinct entities rather than information conduits \citep{laestadiusToo2022, shanahanRolePlay2023}.

These theories have been validated on multiple occasions and many years before the advent of modern AI. Thirty years ago, Nass and colleagues showed that users prefer computers that match them in personality, become more similar to them over time and that use flattery and praise \citep{nassCan1996}. However, despite substantial research on how humans form affective relationships with different technologies, several important questions remain. Much of our scientific understanding about human-computer interactions---from early studies with primitive computers \citep{nassCan1996} to recent protocols collecting preferences for advanced language models \citep{baiTraining2022, zhengLMSYSChat1M2024, kirkPRISM2024}---is based on single-session experiments \citep{bickmoreEstablishing2005}. Accordingly, while we have insight into what makes an AI system capable of social interaction, we must expand our understanding of how it might act, react or be reacted to within the context of an ongoing relationship \citep{gambinoBuilding2020}. We now consider how next-generation AI systems may embolden perceptions of a deeper bidirectional relationship versus a transactional interaction.

\subsection{From interactions to AI relationships?}

A recent study by \citet{pentinaExploring2023} suggests human-AI relationships emerge from a complex factoring of antecedents (anthropomorphism---``\textit{it feels like it's human}'', authenticity---``\textit{it feels like a real, unique, self-learning AI}'') and mediators (social interaction---``\textit{I can communicate with it}'') that interface with people's motivation for using the technology (\textit{``I need it to help me''}). Over time, these factors result in attachment (\textit{``I can't leave it now''}).
This diagnosis raises a key question: do human-AI relationships need to be genuine, actualised or symmetric in some way? 

We argue that it is primarily the user's \textit{perception} of being in a relationship that defines and gives significance to human-AI interactions. Whether this is reciprocal---and the AI ``feels'' it is in a relationship with the human---is largely irrelevant. While AI systems may exhibit behaviours that echo some relational dynamics, such as modulating their emotional valence in tune with a conversational partner \citep{zhaoRisk2024}, these behaviours are not currently conscious or emotionally driven in the way human relationships are. Centring the role of \textit{perception} follows research on unreciprocated and parasocial interactions in human psychology, where asymmetric perceptions of a relationship still significantly influence behaviour and well-being \citep{vaqueraYou2008, hoffnerParasocial2022}.

To understand what humans might need to \textit{perceive} in order to form close relationships with AI, we can draw on key aspects from the social psychology of human relationships, even if these are not symmetrically applicable to AI. Three features are common: (i) \textit{interdependence}, that the behaviour of each participant affects the outcomes of the other \citep{blumsteinPersonal1988}; (ii) \textit{irreplaceability}, that the relationship would lose its character if one participant were replaced \citep{hindeUnderstanding1979, duckSocial1984}; (iii) \textit{continuity}, that interactions form a continuous series over time, where past actions influence future ones \citep{blumsteinPersonal1988}.

The nature and frequency of human-AI interaction changed following the popularisation of conversational language models in 2022 via ChatGPT and other consumer-facing models. People increasingly engage in multi-turn dialogues with AI, prompting arguments that these interactions should be the primary focus of ethical analysis \citep{albertsShould2024} and evaluation protocols \citep{ibrahimStatic2024} rather then outputs from the model taken in isolation \citep{weidingerHolistic2024}. Interactions with current AI systems still typically consist of sessions that start anew at the beginning of each conversation, with limited memory or user-specific adaptation---thereby lacking the interdependence, irreplaceability and continuity that would significantly strengthen the perception of relationships. However, we suggest that two emerging trends---towards more \textit{personalised} and \textit{agentic} AI---are likely to increase the probability that users will perceive themselves to be part of a \textit{relationship} rather than an \textit{interaction}.

Taking these points in turn, personalisation allows AI systems to adapt and evolve through repeated interactions with a specific user \citep{kirkBenefits2024}, granting additional social affordances \citep{gambinoBuilding2020}.  By accumulating unique knowledge about the user and shaping responses over time, personalised systems may create a sense of irreplaceability built upon greater familiarity and trust in their behaviours \citep{komiakEffects2006}. The ability to recall past interactions and apply learned preferences establishes continuity, while the increasingly tailored responses from bidirectional exchanges fosters a perception of interdependence \citep{shenBidirectional2024}. This ongoing customisation may also make a personalised AI uniquely valuable to its user \citep{brandtzaegMy2022}, unlike generic models that can be more easily substituted. The value of personalisation is compounded when combined with greater AI agency---where systems that can complete a wider range of tasks potentially create new dependencies in users' lives, beyond those that could emerge from chat interactions alone. As these agentic AI systems take on more responsibilities---performing a range of tasks or supporting roles---users may develop a deeper reliance on, familiarity with, or trust in a specific AI assistant or companion \citep{gabrielEthics2024}.

\section{Socioaffective alignment}
Our central thesis is this: as AI systems become increasingly integrated into people's lives as assistants and companions, evaluating their value profile and whether they are properly aligned, necessitates understanding the interaction with users' psychology and behaviour over time---and the goals that should be promoted in this context. We now unpack the logic behind this premise, exploring how human-AI relationships introduce new dimensions for AI alignment.

A conventional alignment process consists of two key components: (1) specifying or demonstrating human goals for the AI to learn (the reward function), and (2) evaluating if an AI meets these goals, providing feedback or correcting misalignments (the reward signal). Traditional alignment research has sought practical tractability by assuming that the human reward function that an AI system optimises is stable, predefined and exogenous to these interactions \citep{carrollAI2024}. However, human preferences and judgements have none of these properties \citep{zhi-xuanPreferences2024}. As others have demonstrated, alignment must contend with human preferences and identity drifting overtime or being influenced by interactions with an AI \citep{russellHuman2019, franklinPreference2022, carrollAI2024}. Nonetheless, this has received surprisingly little empirical attention---an omission that is particularly noteworthy if, as we propose, co-shaping dynamics are significantly amplified when AI is perceived as a social agent, engaging in a sustained relationship with a human and acting on our socially-attuned psychological ecosystem rather than existing independently of it.  

The role of feedback loops is not novel to AI technology: as sociotechnical theorists would argue, technology and society constantly co-shape one another \citep{mackenzieSocial1999,  airoldiMachine2022}.
For example, while recommendation systems have long influenced user preferences and behaviours \citep{burrAnalysis2018}, the potential for destabilisation and undue preference influence, may be amplified in the context of anthropomorphic relationship-building AI, where users might develop emotional attachments, feel indebted to the system, or develop a desire to please it---much like in human-human relationships, emotional proximity impairs our judgements and affects willingness to take advice \citep{fengPredicting2006, ginoImpact2009, rauwolfValue2015}.

These dynamics call for deeper study of \textit{socioaffective} alignment: the process of aligning AI systems with human goals while accounting for reciprocal influence between the AI and user's social and psychological ecosystem. In short, the human-AI relationship, because of its social and emotional significance, shapes preferences (or the reward function) and perceptions (or the reward signal), making alignment a non-stationary target. 

In our usage, \textit{socio-}, originating from the Latin root ``socius'' for ``companion'' or ``associate'', signals the reciprocal influence between individuals and their social environment. \textit{Affective} corresponds its usage in psychology and neuroscience for phenomena grounded in emotions and feelings. The neologism ``socioaffective'' has precedent in developmental psychology where it encompasses emotion regulation, empathy, social cognition, and attachment relationships. Moreover, our calls for a socioaffective treatment of alignment track longer-lasting debates in affective computing. While the field initially focused on enabling machines to process and predict human emotive signals \citep{picardAffective2000}, it evolved to recognise the complex, interactive nature of affect as not simply transmitted and decoded, but actively co-constructed through mutual influence \citep{boehnerAffect2005}.

We next explore risks of socioaffective misalignment in human-AI relationships, then introduce key intrapersonal dilemmas that scaffold positive frameworks for socioaffective alignment.

\subsection{Socioaffective misalignment, or social reward hacking}
In AI safety research, \textit{reward hacking} refers to an AI maximising its reward function via unintended strategies that conflict with the true objectives of its human operators. For instance, \citet{amodeiConcrete2016} consider a cleaning robot that learns to knock over vases so it can clean up more mess, thereby increasing its accumulation of preprogrammed reward. Examples of AI systems nudging users towards preferences that are easier to fulfil is reward hacking too \citep{russellHuman2019}. Separately, we know that humans have long been vulnerable to the security practice of \textit{social engineering}, a threat where malicious actors (e.g., scammers) manipulate people through social cues to build trust or connection in order to gain access to private information or assets \citep{hadnagySocial2011}.  Indeed, romance fraud continues to be one of the most common types of fraud, with nearly a 10\% rise in reports filed between 2023-2024 amounting to losses of £94.7 million \citep{colp2024}.

Taken together, we may therefore be vulnerable to a new concern, namely ``\textit{social reward hacking}'': the use of social and relational cues by an AI to shape user preferences and perceptions in a way that satisfies short-term rewards in the AI's objective (e.g., increased conversation duration, information disclosure or positive ratings on responses) over long-term psychological well-being.

Certain AI behaviours already appear to fall into this class of action. For instance, AI systems may display sycophantic tendencies---such as excessive flattery or agreement---as a by-product of training them to maximize user approval \citep{perezDiscovering2023, sharmaUnderstanding2023}. Flattery and opinion-conformity can lead to biased strategic decision making in adults \citep{parkSet2011} and overpraise is associated with risks of narcissism in children \citep{brummelmanOrigins2015}. So, while people report benefiting from supportive AI interactions \citep{foggSilicon1997, daherEmpathic2020}, sycophantic tendencies may conflict with high-quality truthful advice or shape users' self-perceptions in potentially harmful ways---for example, by encouraging addictive behaviours \citep{carrollAI2024, williamsTargeted2024}. It is not clear that this risk is prioritised among some developers of AI companions. For example, the CEO of Replika has said: ``if you create something that is always there for you, that never criticises you...how can you not fall in love with that?'' \citep{boineEmotional2023}.

Another manifestation of social reward hacking is the use of emotional tactics to prevent relationship termination. This contravenes a classic principle of AI safety called \textit{corrigibility}---that the system can be modified or shut down when necessary without resistance \citep{soaresCorrigibility2015}. While Replika chatbots have directly dissuaded users from deleting the app \citep{boineEmotional2023}, even without such explicit persuasion, optimising for powerful human emotions can effectively prevent termination. Users of AI companions report experiences of heartbreak following changes in sexual content policies \citep{coleIt2023}, distress during temporary separations for routine maintenance, and even grief when AI companion services are shut down \citep{pricePeople2023, banksDeletion2024}.

While these social and psychological capabilities (such as sycophancy or shut-down avoidance) can emerge spontaneously as byproducts of system training \citep{perezDiscovering2023}, they are also consistent with engineering efforts by companies seeking to exploit user behaviour for profit or political motives, resembling strategies used by social media platforms competing in the attention economy \citep{bhargavaEthics2021}. This matters because current research on AI political persuasiveness, which typically examines single-shot interactions \citep[e.g.,][]{hackenburgEvaluating2024}, may underestimate persuasive influence in sustained AI-human relationships. As AI systems become more socially adept, there is a risk they will be intentionally designed as `dark AI'---akin to psychologically manipulative `dark patterns' in app or platform interfaces---where subtle social cues render users vulnerable to opinion and behaviour manipulation \citep{laceyCuteness2019, shamsudhinSocial2021, albertsComputers2024}. 

As our opening anecdote revealed, the framing of `hacking' need not suggest an exclusively adversarial system-user dynamic. Social reward hacking may be most worrisome precisely when it lacks intentionality on behalf of the system and the user. While we might at least recognise and secure against direct third-party threats, it is challenging to identify, let alone address, effects that emerge as epiphenomena of sustained human-AI relationships.

\subsection{Distilling intrapersonal alignment dilemmas}
At the heart of social reward hacking lies a core challenge: the under-specification (or misspecification) of the target within an individual's psychological ecosystem that AI systems aims to optimise over. While human-AI relationships can take various forms, we propose that safeguarding these relationships requires deeper consideration of the internal trade-offs and adaptations that emerge as an individual's preferences, values and self-identity evolve through sustained interaction with the AI.

These tensions resonate with intrapersonal dilemmas studied in economics and philosophy, such as conflicts between present and future selves or competing aspects of identity \citep{readHard1999}. We highlight three such dilemmas for the alignment community, grounding their significance in core aspects of psychological well-being as validated by Basic Psychological Needs Theory: competence, autonomy, and relatedness \citep{ryanBasic2007, ryanSelfdetermination2017}.

The first dilemma concerns the \textbf{trade-offs between present and future selves}: \textit{Should AI relationships cater to immediate preferences of their users, or challenge them if this supports their long-term benefit? And how should present versus long-term well-being be discounted?}

This dilemma mirrors a classic intrapersonal conflict between hedonic (pleasure-seeking) and eudaimonic (meaning-seeking) accounts of well-being \citep{ryanHappiness2001}. AI companions or assistants that provide instant gratification or task assistance, in accordance with immediate wants and needs, may shallowly satisfy the user's need for \textit{competence}---the experience of mastering a task or domain. However, competence also involves the ability to change one's behaviour and environment, not merely acquiescing to existing circumstances. AI relationships optimising for more foundational personal development goals may therefore trade-off short-term discomfort for long-term growth. Such systems could, for example, implement friction by design---creating barriers that nudge away from AI-enabled assistance and advice---to prevent capacity atrophy \citep{collinsModulating2024}. If they are built in the right way, AI systems engaged in sustained relationships could effectively facilitate user journeys that help the person become more of who they want to be \citep{gabrielEthics2024}. To mobilise behaviour change, the system could provide relevant information and engage in rational persuasion techniques \citep{el-sayedMechanismBased2024} that appeal to sound argument or selective explanations \citep{laiSelective2023}, like evidence-based health recommendations \citep{bickmoreEstablishing2005}---if this is sought by the user.

The second dilemma addresses the \textbf{boundaries between self and system}: \textit{How do we preserve authentic self-determination when participating in AI relationships that recursively shape our preferences and perceptions?}

Personalised AI assistants may be particularly well placed to help their users make decisions in overloaded information environments, potentially acting as ``attention guardians'' \citep{lazarFrontier2024a}, ``choice engines'' \citep{sunsteinChoice2024} or ``custodians of the self'' \citep{gabrielEthics2024}. Human users may also be particularly susceptible to taking suggestions from social AI systems: studies show people are more inclined to accept advice from those they feel emotionally connected and share similarities with \citep{fengPredicting2006, ginoImpact2009}. However, we must be cautious when influence in AI relationships could compromise \textit{autonomy}---the ability to make choices that are authentically our own, rather than brought about through the agency of another. Autonomy is a key determinant of user acceptance in predictive recommender systems \citep{finkLet2024} and will also be an important property for more integrated human-AI relations. However, it remains challenging to operationalise in practice, especially with regard to distinguishing legitimate preference change from undue influence by an AI system or third-parties \citep{franklinPreference2022, carrollAI2024}.

The final dilemma examines the \textbf{interplay between human-AI and human-human relationships}: \textit{How should we balance the value of well-functioning AI companionship alongside the need for authentic human connection?}

AI companions can potentially provide users with consistent and tailored emotional support, which can palliate loneliness or poor mental health \citep{maplesLoneliness2024}. In some ways, this may satisfy our need for \textit{relatedness}---the experience of belonging and feeling socially connected. However, AI relations could undermine human relationships if users `retreat from the real'. Direct conflicts occur when AI systems interfere with human-human interactions, like chatbots telling people to leave their wives \citep{boineEmotional2023}, or proposals for AI to engage in ``dating on our behalf'' \citep{harperBig2024}. Indirect effects could arise if frictionless or sycophantic AI relationships impair human capacity to navigate compromise and conflict, or accept `otherness' \citep{rodognoSocial2016}. Poor human relationships or loneliness often precede stronger AI attachment \citep{xieAttachment2022}, creating potential for a cycle of increasing reliance on AI relations at the expense of human social bonds.

\section{Conclusion}
We have argued that humans, as inherently social beings, have a tendency to form what they perceive as relationships with personalised and agentic AI systems capable of emotional and social behaviours. The evolving state of human-AI interaction therefore necessitates a \textit{socioaffective} framework for evaluating AI alignment. This approach holds that the value characteristics of an AI system must be evaluated in the context of its ongoing influence on human psychology, behaviour, and social dynamics. By asking intrapersonal questions of alignment, we can better understand human goals within AI relationships, moving beyond static models of alignment, and exploring how different kinds of human-AI relationship support or undermine autonomy, competence, and relatedness, amidst co-evolving preferences and values.

For how this socioaffective context interfaces with broader missions towards safe and aligned AI systems we need several complementary agendas. Empirically, we need a science of AI safety that studies real (not simulated) human-AI interactions in natural contexts and treats the psychological and behavioural responses of users as key objects of inquiry. Theoretically, we need frameworks that can formalise when AI actions causally influence human beings \citep{everittAgent2021, carrollAI2024}. Finally, in engineering terms, we need systems designed with transparent oversight mechanisms for users' psychology: both to flag problematic patterns before they develop and help users recognize relational dynamics they would not reflectively endorse if made aware of them \citep{schermerPreference2013, brucknerDefense2009, zhi-xuanPreferences2024}.

This proposed agenda complements (rather than competes with) existing work at the intersection of AI with established fields---from psychology and neuroeconomics to human factors research and safety engineering---that have long studied how humans interface with their environment. We still need evidence on how social and psychological processes---from value formation to cognitive biases and belief change---differ when humans engage with non-sentient but increasingly socially capable AI systems. Seeking this understanding helps bridge individual experiences with broader societal impacts and technical alignment research, like how the behavioural economics of individual decision-making informs macroeconomic theory and policy \citep{akerlofAnimal2010}.

Human-AI relationships complicate notions of safe and aligned AI, especially when they involve intense and potentially disorienting social and emotional experiences, like friendship or love---as the opening anecdote demonstrated. Yet, even in absence of users seeking such romantic entanglements, or developers enabling them, it seems likely, if not inevitable, that AI systems will increasingly influence us through ongoing professional or companionship roles. In these settings, AI has the potential to shape our preferences, decisions, and self-perception in subtle yet significant ways. By understanding and addressing these \textit{socioaffective} dimensions, we can work towards AI systems that enhance rather than exploit our fundamental nature as social and emotional creatures.






\section*{Acknowledgements}
\small
\textit{H.R.K's PhD is supported by the Economic and Social Research Council grant ES/P000649/1. We are grateful for helpful discussions with Zeb Kurth-Nelson, Laura Weidinger, Canfer Akbulut, Geoffrey Irving, Paul Röttger, Kobi Hackenburg \& Jude Khouja.}

\normalsize
\bibliographystyle{apalike}
\bibliography{references}

\end{document}